\documentclass{appolb}
\usepackage{epsfig}

\begin{document}

\newcommand{\drbar}{{\overline{DR}}}
\newcommand{\msbar}{{\overline{MS}}}
\newcommand{\GeV}{{\rm GeV}}

\newcommand{\stau}{\tilde{\tau}}
\newcommand{\snt}{{\tilde{\nu}_\tau}}
\newcommand{\ur}{\tilde{u}_R}
\newcommand{\ul}{\tilde{u}_L}
\newcommand{\dr}{\tilde{d}_R}
\newcommand{\dl}{\tilde{d}_L}
\newcommand{\st}{\tilde{t}}
\newcommand{\sbot}{\tilde{b}}
\newcommand{\sg}{\tilde{g}}
\newcommand{\nt}{\tilde{\chi}^0}
\newcommand{\cp}{\tilde{\chi}^+}
\newcommand{\cm}{\tilde{\chi}^-}
\newcommand{\cx}{\tilde{\chi}}
\newcommand{\ser}{\tilde{e}_R}
\newcommand{\sel}{\tilde{e}_L}
\newcommand{\sne}{\tilde{\nu}_e}

\pagestyle{plain}
\newcount\eLiNe\eLiNe=\inputlineno\advance\eLiNe by -1
\title{On the SPA Convention and Project%
\thanks{Talk given at the International Workshop on Photon Linear Colliders,
  Kazimierz, Poland, 5-8 September 2005}%
}
\author{Jan KALINOWSKI\thanks{Supported by the Polish KBN Grant  2 P03B 040 24
    (for years 2003-2005)}%
\address{Institute of Theoretical Physics, Warsaw University,
Ho\.za 69, 00-681~Warsaw, Poland}}
\maketitle

\begin{abstract}
  Reconstruction of the fundamental supersymmetric theory and its breaking
  mechanism will require high-precision tools. Here a brief introduction to
  SPA, the Supersymmetry Parameter Analysis (SPA) Convention and Project, is  
  presented which is based on a consistent set of conventions
  and input parameters. 
\end{abstract}

\section{Introduction}
At future colliders, the LHC and the ILC, experiments can be performed in the
supersymmetric particle sector 
with very high precision --
experimental accuracies are expected at the per-cent down to the per-mille
level \cite{R2,R3A,R2B}.  This should be matched from the theoretical
side~\cite{Kal}. Therefore a well defined
theoretical framework is necessary for the calculational schemes in
perturbation theory as well as for the input parameters. Motivated by the
experience   in analyzing data  
at the former $e^+e^-$ colliders LEP and
SLC, the SPA  
Convention and Project~\cite{SPA} has been proposed. It
provides \\
\phantom{mm}$\bullet$ a convention for calculating masses, mixings, decay
widths and
production cross sections,\\
\phantom{mm}$\bullet$ a program repository of codes (RGE, spectrum
calculators, fitting routines, event generators etc.)
that will be expanded continuously in the future,\\
\phantom{mm}$\bullet$ a list of future tasks on both the theoretical and the
experimental sides needed before data from future experiments could be
evaluated at the desired level 
of accuracy,\\
\phantom{mm}$\bullet$ a SUSY reference point SPS1a$'$ as a general setup for
testing these tools in practice.\\
Combining the experimental 
information from LHC and ILC  
will provide a  
high-precision picture of supersymmetry at the
TeV scale \cite{Allanach:2004ed} which subsequently may lead to 
the reconstruction of the fundamental supersymmetric
theory at a high scale and the mechanism of
supersymmetry breaking \cite{atGUT}.   

The SPA Convention and Project is a joint inter-regional experimental and
theoretical  effort. The current status of the project 
is documented on  the routinely updated web-page\\
\phantom{mmmmmmmmmmmmmm}{\bf  http://spa.desy.de/spa/  }

\section{SPA CONVENTION}
Building on vast  experience in SUSY calculations and data simulations and 
analyses, 
the SPA Convention consists of the following propositions:
\begin{itemize}
   \item  The masses of the SUSY particles and Higgs bosons are defined as pole
     masses.
   \item All SUSY Lagrangian parameters, mass parameters and couplings,
     including 
     $\tan\beta$, are given in the $\drbar$ scheme and defined at the scale
     $\tilde M =$ 1 TeV.\\[-8mm]
   \item Gaugino/higgsino and scalar mass matrices, rotation matrices and the
     corresponding angles are defined in the $\drbar$ scheme at $\tilde M$,
     except for the Higgs system in which the mixing matrix is defined in the
     on-shell scheme, the scale parameter chosen as the light Higgs mass.
   \item The Standard Model input parameters of the gauge sector are chosen as
     $G_F$, $\alpha$, $M_Z$ and $\alpha_s^{\msbar}(M_Z)$. All lepton masses are
     defined on-shell. The $t$ quark mass is defined on-shell; the $b,\, c$
     quark masses are introduced in $\msbar$ at the scale of the
     masses themselves while taken at a renormalization scale of 2 GeV for
     the light $u,\, d,\, s$ quarks.
   \item Decay widths, branching ratios and production cross sections are
     calculated for the set of parameters specified above.
\end{itemize}

The $\drbar$  scheme, 
based on dimensional reduction and  modified
minimal subtraction,  is designed
to preserve supersymmetry and it is technically very convenient. 
Recently it has been shown that it 
can be formulated in a mathematically consistent way  \cite{Stockinger:2005gx}
and  to comply with the 
factorization theorem, see~\cite{37b}. The physical on-shell masses are
introduced in the decay widths and 
production cross sections such that the phase space is treated
in the observables closest to experimental on-shell kinematics.

\section{Program repository}

To use the highly developed $\msbar$ 
infrastructure for proton colliders a repository contains   
the translation tools between the $\drbar$ and $\msbar$ schemes, 
as well as the on-shell renormalization schemes. 
The responsibility for developing codes and maintaining them up to the current 
theoretical state-of-the-art
precision rests with the authors. The SLHA
\cite{Skands:2003cj} convention is recommended for communication between
the codes. 
The repository contains links to codes grouped in several categories:
\begin{enumerate} 
  \item Scheme translation tools for 
    definitions and relations between on-shell, $\drbar$ and
    $\msbar$ parameters.
  \item Spectrum calculators for 
    transition from the Lagrangian parameters to a
    basis of physical particle masses and the related mixing matrices. 
  \item Calculation of other observables: 
decay tables, cross sections, low-energy observables etc.
    \item Cosmological and astrophysical aspects: cold dark matter
      relics, cross sections for CDM particle searches, astrophysical
      cross-sections in the SUSY context etc.
  \item Event generators that generate event samples for SUSY and background
    processes in realistic collider environments.
  \item Analysis programs to extract the
    Lagrangian parameters from experimental data by means of global
    analyses.    
  \item RGE programs to connect the
low-energy effective Lagrangian parameters to the high-scale
    where the model is supposed to be matched to a more fundamental
    theory. 
  \item Auxiliary programs and libraries:    
    structure functions, beamstrahlung, numerical methods, SM
    backgrounds etc.
  \end{enumerate}

\section{Testing the SPA: Ref.\ Point SPS1a$'$ }

The SPA Convention and Project is a very ambitious and extended experimental
and theoretical effort.  It is set up to cover general SUSY scenarios.
However, to perform first checks of its internal consistency and to explore
the potential of such coherent data analyses a MSSM Reference Point SPS1a$'$
has been proposed as a testing ground. Of course, in future the SPA has to be
tested in more extended MSSM as well as more complicated scenarios.

The roots defining the Point SPS1a$'$ are the mSUGRA
parameters [in the conventional notation for cMSSM]
in the set 
\begin{eqnarray*}
  \fbox{$%
    \begin{array}{lcrlcr}
      M_{1/2} &=& 250 \;\mbox{GeV}  \;\;  &  \mbox{sign}(\mu)      &=& +1
      \\  
      M_0     &=&  70 \;\mbox{GeV}  \;\;  &  \tan\beta (\tilde M)  &=& 10
      \\ 
      A_0     &=&-300 \;\mbox{GeV}  \;\;  &                        & &         
    \end{array}  $} 
  \label{param}
\end{eqnarray*}
where the universal gaugino mass $M_{1/2}$,
the scalar mass $M_0$ and the trilinear coupling $A_0$ [Yukawa couplings
factored out], are 
defined at the GUT scale $M_U$. The point is
close to the original Snowmass point SPS1a~\cite{Allanach:2002nj} and   
to point B$'$ of
\cite{Battaglia:2003ab}.
With the SM input parameters given explicitly in the SPA document
\cite{SPA}, 
extrapolation of the above parameters 
down to the $\tilde{M} = $ 1 TeV scale 
generates the MSSM Lagrangian parameters as shown in the left  
part of Table~\ref{tab:par}. Here  
the RGE part of the program
{\ttfamily SPheno} \cite{Porod:2003um} has been used  
(for internal or external
comparison, other codes can equally be used).

\renewcommand{\arraystretch}{1.2}
\begin{table}[t] \small
\begin{minipage}{8cm}
$ \begin{array}{|c|c||c|c|}
\hline
 g'   &   0.3636  &  M_1       & 103.3  \\
 g    &   0.6479  &  M_2       & 193.2  \\
 g_s  &   1.0844  &  M_3       & 571.7  \\ \hline
 Y_\tau & 0.1034  & A_\tau     & -445.2 \\
 Y_t  &   0.8678  & A_t        & -565.1 \\
 Y_b  &   0.1354  & A_b        & -943.4 \\ \hline
 \mu  &    396.0  & \tan\beta  & 10.0   \\ \hline
 M^2_{H_1}  & 2.553 \cdot 10^4 & 
 M^2_{H_2}  &-14.31 \cdot 10^4  \\ \hline  
 M^2_{L_1}  & 3.278 \cdot 10^4 &  M^2_{L_3}  & 3.214 \cdot 10^4 \\
 M^2_{E_1}  & 1.338 \cdot 10^4 &  M^2_{E_3}  & 1.210 \cdot 10^4 \\
 M^2_{Q_1}  & 27.64 \cdot 10^4 &  M^2_{Q_3}  & 22.22 \cdot 10^4 \\
 M^2_{U_1}  & 25.73 \cdot 10^4 &  M^2_{U_3}  & 15.04 \cdot 10^4 \\
 M^2_{D_1}  & 25.50 \cdot 10^4 &  M^2_{D_3}  & 25.09 \cdot 10^4 \\ \hline
\end{array} $
\end{minipage}\begin{minipage}{7cm}
$\begin{array}{|c|c||c|c|}
\hline
  h^0    &  116.0 & \stau_1 &  107.9 \\
  H^0    &  425.0 & \stau_2 &  194.9 \\
  A^0    &  424.9 & \snt    &  170.5 \\ \cline{3-4}
  H^+    &  432.7 & \ur     &  547.2 \\ \cline{1-2}
  \nt_1  &   97.7 & \ul     &  564.7 \\
  \nt_2  &  183.9 & \dr     &  546.9 \\
  \nt_3  &  400.5 & \dl     &  570.1 \\ \cline{3-4}
  \nt_4  &  413.9 & \st_1   &  366.5 \\
  \cp_1  &  183.7 & \st_2   &  585.5 \\
  \cp_2  &  415.4 & \sbot_1 &  506.3 \\ \cline{1-2}\cline{1-2}
  \ser   &  125.3 & \sbot_2 &  545.7 \\ \cline{3-4}
  \sel   &  189.9 & \sg     &  607.1 \\
  \sne   &  172.5 &         &        \\
  \hline
\end{array}    $ 
\end{minipage}
\caption{\small \it {\rm Left:} 
The $\drbar$ SUSY Lagrangian parameters 
  at the scale $\tilde M=1$~{\rm{TeV}} 
  in SPS1a$'$ from \cite{Porod:2003um}. In addition, 
  gauge and Yukawa couplings at this scale are given in the 
  $\drbar$ scheme. {\rm Right:} Mass spectrum of
  supersymmetric particles \cite{Porod:2003um} and
  Higgs bosons \cite{Heinemeyer:1998yj} in the reference point SPS1a$'$.
  The masses in the second generation coincide
  with the first generation.  [Mass unit in {\rm GeV}]}
\label{tab:par}
\end{table}

The physical [pole] masses of the supersymmetric particles are collected
in the right part of Table~\ref{tab:par}.
The connection between the Lagrangian parameters and the physical pole
masses can presently be controlled at the 1-loop level for the masses of
the SUSY particles, and at the 2-loop level for the Higgs masses.
The QCD effects on the heavy quark masses are accounted for to 2-loop
accuracy. 

\begin{figure}[htb]
  \begin{center}
  \includegraphics[height=5cm, width=6cm]{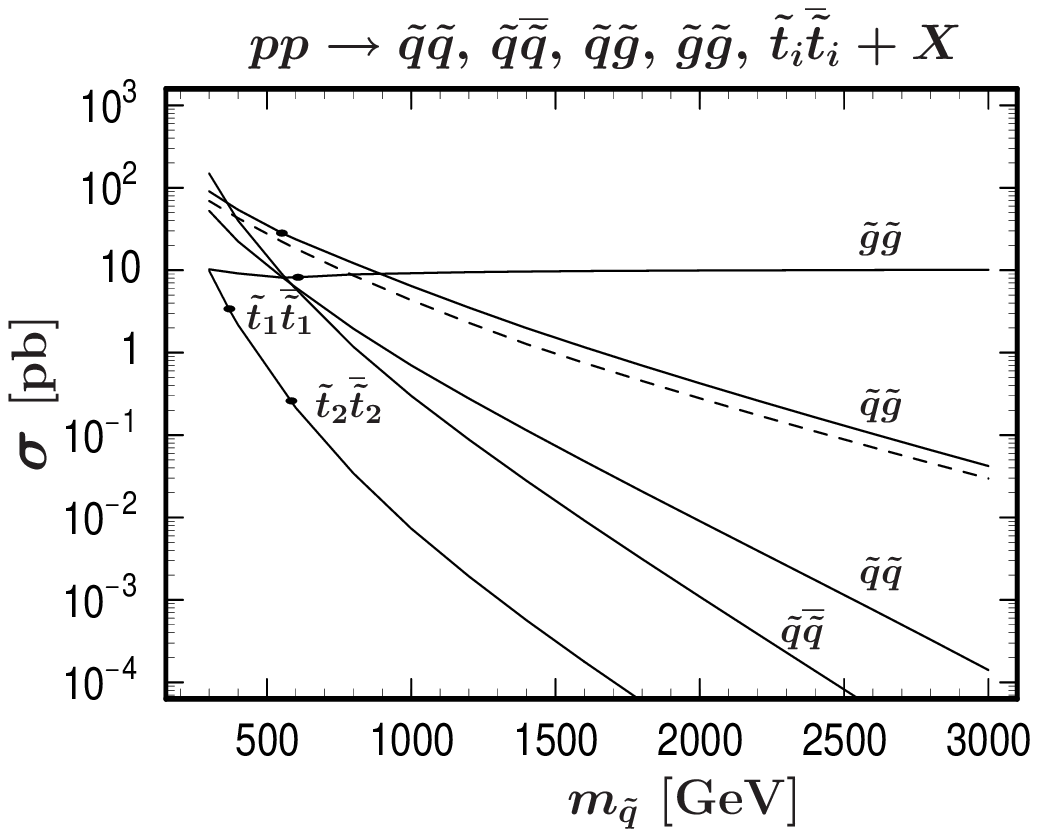}
  \includegraphics[height=5cm, width=6cm]{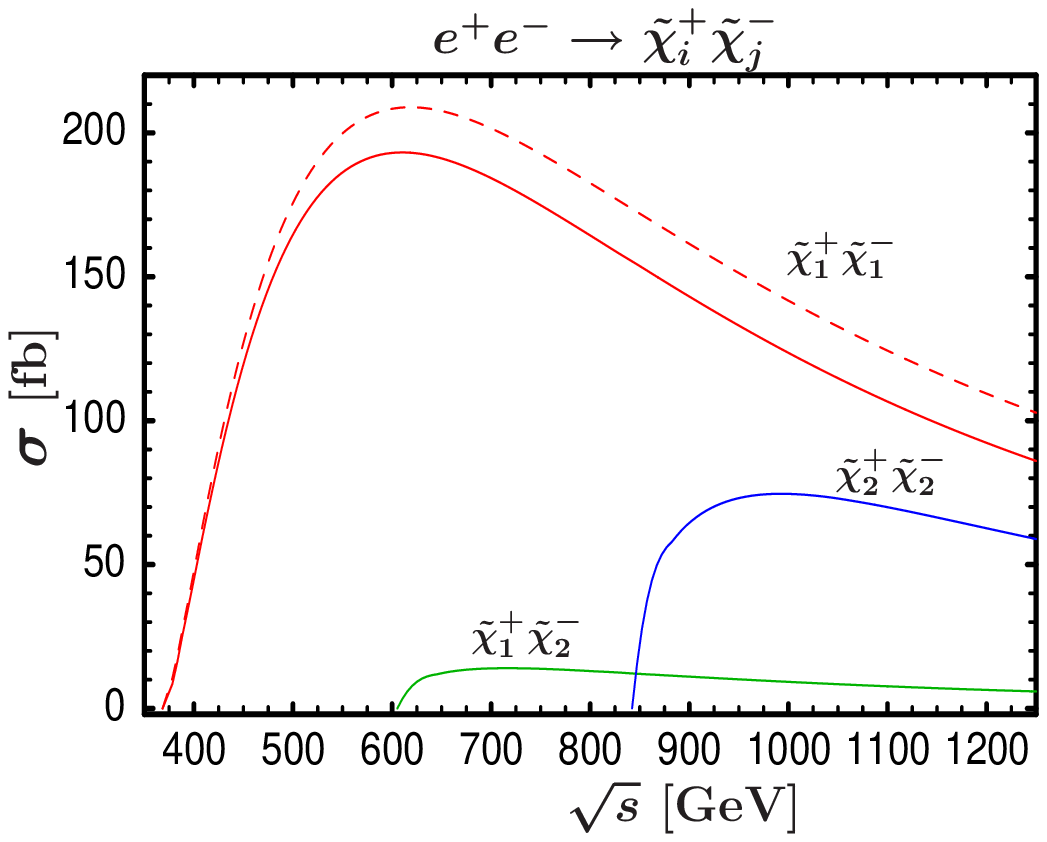}
\end{center}
  \caption[]{\it {\rm Left:} Total cross-sections for squark and gluino 
    pair production at the LHC~\cite{prospino}
as a function of squark mass keeping the 
    gluino mass fixed. {\rm Right:} Total cross section sections for chargino
      pair production in $e^+e^-$ annihilation.
      The Born cross sections (broken lines) are shown for a few channels.}
\label{fig:xsec}
\end{figure}

For illustration, the left panel of Figure~\ref{fig:xsec}  displays 
cross sections for the production of squarks and gluinos at the 
LHC 
as  functions of the squark mass  crossing the point SPS1a$'$,    
while the right panel shows the 
production cross section of pairs of charginos for the point SPS1a$'$ 
at the ILC as a function of the collider energy.

To perform experimental simulations, the branching ratios of the
decay modes are crucial. The SPA Document and the SPA web page 
provide results of  calculations using 
{\ttfamily FeynHiggs}~\cite{Heinemeyer:1998yj} 
and {\ttfamily SDECAY}~\cite{sdecay}.   

If SPS1a$'$, or a SUSY scenario with 
mass scales similar to this point, 
is realized in nature, a plethora of
interesting channels can be exploited to extract the basic
supersymmetry parameters when combining experimental information from
mass distributions at LHC with measurements of decay
spectra and threshold excitation curves at an $e^+e^-$ collider with
energy up to 1 TeV {\cite{Allanach:2004ed}}. 
From the simulated experimental errors the data
analysis performed coherently for the two machines gives rise to a
very precise picture of the supersymmetric particle spectrum as
demonstrated in the left part of Table~\ref{tab:massesA}.

\renewcommand{\arraystretch}{1.2}
\begin{table}[h] \footnotesize
\begin{minipage}{9.5cm} $
  \begin{array}{|c|c||c|c||c|}
    \hline 
    \ \mbox{Particle} \ &
    \ \ \mbox{Mass}\ \  & \ \mbox{LHC}\ & \ \mbox{ILC}\ 
                        & \ \mbox{LHC+ILC}\ \\ 
    \hline\hline
    h^0                 & 116.9 & 0.25 & 0.05 & 0.05 \\
    H^0                 & 425.0 &      & 1.5  & 1.5  \\
    \hline 
    \tilde{\chi}^0_1    &  97.7 & 4.8  & 0.05 & 0.05 \\
    \tilde{\chi}^0_2    & 183.9 & 4.7  & 1.2  & 0.08 \\
    \tilde{\chi}^0_4    & 413.9 & 5.1  & 3-5  & 2.5  \\
    \tilde{\chi}^\pm_1  & 183.7 &      & 0.55 & 0.55 \\ \hline 
    \tilde{e}_R         & 125.3 & 4.8  & 0.05 & 0.05 \\
    \tilde{e}_L         & 189.9 & 5.0  & 0.18 & 0.18 \\
    \tilde{\tau}_1      & 107.9 & 5-8  & 0.24 & 0.24 \\ \hline
    \tilde{q}_R         & 547.2 & 7-12 & -    & 5-11 \\
    \tilde{q}_L         & 564.7 & 8.7  & -    & 4.9  \\
    \tilde{t}_1         & 366.5 &      & 1.9  & 1.9  \\
    \tilde{b}_1         & 506.3 & 7.5  & -    & 5.7  \\ \hline
    \tilde{g}           & 607.1 & 8.0  & -    & 6.5  \\ \hline
  \end{array}$ \\
\end{minipage}\begin{minipage}{5cm}
  \begin{tabular}{|c|c|}
    \hline
    Param.\  & Error  \\
    \hline\hline
    $M_1$                  &    $ 0.1  $  \\
    $M_2$                  &    $ 0.1  $  \\
    $M_3$                  &    $ 7.8  $  \\
    $\mu$                  &    $ 1.1  $  \\ \hline
    $M_{\tilde{e}_L}$      &    $ 0.2   $  \\
    $M_{\tilde{e}_R}$      &    $ 0.4   $  \\
    $M_{\tilde{\tau}_L}$   &    $ 1.2   $   \\ \hline
    $M_{\tilde{u}_L}$      &    $ 5.2   $   \\
    $M_{\tilde{u}_R}$      &    $ 17.3 \;\;  $   \\
    $M_{\tilde{t}_L}$      &    $ 4.9   $    \\ \hline
    $m_{\mathrm{A}}$       &    $ 0.8  $   \\
    $A_{\mathrm{t}}$       &    $ 24.6 \;\,  $   \\
    $\tan\beta$            &    $ 0.3 $  \\ \hline
  \end{tabular}
\end{minipage}
\caption{ \small For the point SPS1a$'$ -- \mbox{Left:} 
\it Accuracies for representative mass measurements
    of SUSY particles in individual LHC, ILC and 
    coherent LHC+ILC analyses.
    $\tilde q_R$ and $\tilde q_L$ represent 
    $q=u,d,c,s$. \mbox{\rm Right:} Excerpt of 
    extracted SUSY Lagrangian mass and Higgs parameters 
  at the supersymmetry scale $\tilde{M} =$ 1 TeV.  Masses in {\rm GeV}.} 
\label{tab:massesA}
\end{table}

In addition to  evaluating the experimental
observables channel by channel, global analysis programs 
have become available \cite{Lafaye:2004cn} in 
which the whole set of data, masses, cross sections, branching ratios etc, 
is exploited coherently to extract the
Lagrangian parameters in the optimal way after including the available
radiative corrections. The present quality
of such an analysis can be judged from the results 
shown in the right part of  Table~\ref{tab:massesA}, 
where only experimental errors are included.

\begin{figure}[htb]
{\small $1/M_i$~[GeV$^{-1}]$ ~~~~~~~ ~~~~~~~~~ ~~~~~~~
 $M^2_{\tilde j}$~[$10^3$ GeV$^2$]}
\centering
  \includegraphics[height=5cm, width=6cm]{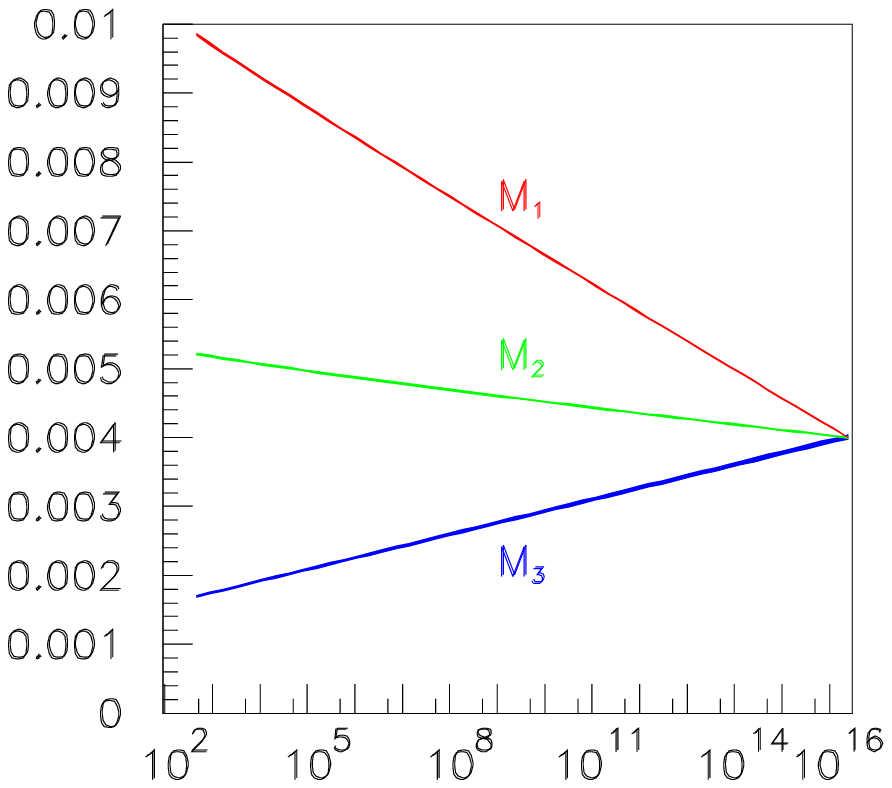}
  \includegraphics[height=5cm, width=6cm]{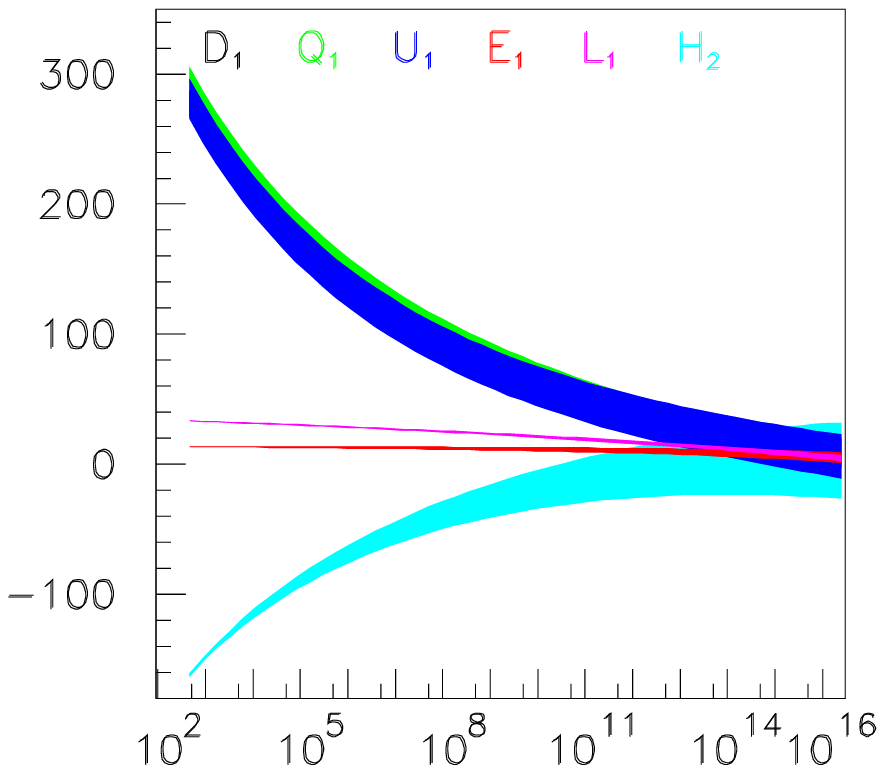}\\
$Q$~[GeV] ~~~~~~~~~~~  ~~~~~~~~~~~~~~~~ $Q$~[GeV]
\caption{\small \it 
  Running of the gaugino and scalar mass parameters in SPS1a$'$ 
  \cite{Porod:2003um}.
  Only experimental errors are taken into account; theoretical errors are
  assumed to be reduced to the same size in the future.}
\label{fig:running}
\end{figure}

With the parameters extracted at the scale $\tilde{M}$, the reconstruction of
the fundamental supersymmetric theory and the related microscopic picture of
the mechanism breaking supersymmetry can be attempted.  In the bottom-up
extrapolation \cite{atGUT} from $\tilde{M}$ to the GUT/Planck scale by the
renormalization group evolution for all parameters the experimental
information is exploited to the maximum extent possible.  The evolution of the
gaugino and scalar mass parameters for SPS1a$'$ are presented in
Figure~\ref{fig:running}. While the reconstruction of the high-scale
parameters in the gaugino/higgsino and slepton sectors is very precise, the
picture of the colored scalar and Higgs sectors is still coarse and strong
experimental efforts on improving mass and trilinear coupling
measurements should be made to refine it considerably.

On the other hand, if the 
structure of the theory at the high scale was known {\it a priori} and
only the experimental determination of the high-scale parameters was  
lacking, then the 
top-down approach would lead to a very precise parametric picture at the
high scale. This is apparent from the fit of the mSUGRA parameters in
SPS1a$'$ displayed in Table~\ref{tab:univ_params}. 

\renewcommand{\arraystretch}{1.2}
\begin{table}[htb]
\begin{center} \footnotesize
  \begin{tabular}{|c||c|c|}
    \hline
    Parameter       &  SPS1a$'$ value           & Experimental error \\ 
    \hline
    $M_\frac{1}{2}$ & {\phantom{-}}250 GeV      &  0.2  GeV    \\
    $M_0$           & {\phantom{-0}}70 GeV      &  0.2  GeV     \\
    $A_0$           & -300 GeV                  &  13.0 GeV $\;$  \\  
    $\mu$           &  396.0 GeV                &  0.3 GeV    \\
    $\tan\beta $    &  10                       &  0.3    \\  \hline
  \end{tabular}
\end{center}
\caption{\small \it Comparison of the ideal parameters with the
  experimental expectations in the top down approach.} 
\label{tab:univ_params}
\end{table}

\section{Future developments}

The results for SPS1a$'$ presented here are based
on preliminary experimental simulations 
and extrapolations from earlier analyses for SPS1a
and other reference points as a 
substitute of missing information necessary for a first comprehensive
test of all aspects of the SPA Project. 
Althought current SPA studies are very encouraging, 
much additional work both on 
the theoretical as well as on the experimental side will be needed to achieve
the SPA goals.  In particular
\begin{itemize}
\item The present level of theoretical calculations still does not match the
  expected experimental precision, particularly in coherent LHC+ILC analyses.
  New efforts in higher-order SUSY calculations are necessary for the
  interpretation of experimental analyses.
\item There is no complete proof that $\drbar$ scheme preserves supersymmetry
  and gauge invariance in all cases.  As the precision of SUSY calculations is
  pushed to higher orders, the SPA Project also requires an improved
  understanding of the $\drbar$ scheme.  Recently it has been shown that for
  massive final state particles spurious density functions for the $(4 - D)$
  gluon components have to be introduced to comply with the factorization
  theorem~\cite{37b} which opens a way to formulating an efficient combination
  of the most attractive elements of $\drbar$ and $\msbar$ schemes in
  describing hadronic processes.
\item A limited set of observables included in experimental
  analyses by no means exhausts the opportunities which data at LHC and at
  ILC are expected to provide in the future.  Likewise, in most analyses
  errors are purely experimental and do not include the theoretical
  counterpart which must be improved considerably before matching the
  experimental standards.  This is particularly important for the coherent
  combination of future data obtained at LHC and ILC.  Feedback and coherently
  combined analyses, which will greatly benefit from a concurrent running of
  both colliders, are indispensable for a meaningful reconstruction of the
  underlying theory~\cite{R2B}.
\item Astrophysical data play an increasingly important role in confronting
  supersymmetry with experiments. Models with R-parity conservation predict a
  stable weakly interacting, massive particle. In this case on the one hand
  their relic abundance imposes crucial limits on supersymmetric scenarios and
  specific requirements on the accuracies must be achieved when the CDM
  particle is studied in high-energy physics laboratory
  experiments~\cite{puk}. 
  On the other, predictions based on the comprehensive parameter analysis of
  high-energy experiments determine the cross sections for astrophysical
  scattering experiments by which the nature of the cold dark matter particles
  can be established. For example, a study of the SPS1a point at LHC, based on
  very large statistics \cite{tovey}, indicates that the relic density can be
  determined to $\sim 6$\% for the SPS1a$'$ scenario.  At the ILC a precision
  of $\sim 1.5$\% should be achievable.  
\item    
The parameter set SPS1a$'$ chosen  for a first study  
   provides a benchmark  for
   developing and testing the tools needed for a successful analysis
   of future SUSY data.  However, neither this specific point nor
   the MSSM itself may be the correct model for low-scale SUSY.
While versions of mSUGRA and of
  gaugino mediation have also been analyzed in some detail, 
the analyses have to be
  extended systematically to other possibilities.   In particular,  
   CP violation, R-parity violation, flavor violation, NMSSM and
   extended gauge groups are among scenarios which might be realised 
   in the SUSY sector. The SPA conventions are 
   general enough to cover all these
   scenarios.

\end{itemize}   

\section{Summary}
The SPA Project, a joint
theoretical and experimental effort, aims at providing\\[1mm]
\phantom{mm}$\bullet$ a well defined framework for SUSY calculations and
    data analyses,\\
\phantom{mm}$\bullet$ all necessary theoretical and computational tools,\\
\phantom{mm}$\bullet$ a testground scenario SPS1a$'$,\\
\phantom{mm}$\bullet$ a platform for future extensions and
developments.\\[1mm] 
First results for the reference point SPA1a$'$ are very encouraging, however
it is clear that 
much work is still needed on the experimental and
theoretical sides to achieve the desired
level of accuracy. The SPA Project is a dynamical system expected to
evolve continuously and to encompass more general supersymmetry
scanarios.\\[3mm]

\noindent {\bf Acknowledgments:}\\[1mm]  I would like to thank all 
the SPA members for their efforts to make the SPA Projects 
a success.  Special thanks go to Peter Zerwas and Uli Martyn.\\    I am also
deeply indebted to Prof.\ Wojciech Marczy\'nski, MD,  Dr. Tomasz Skrok,
MD, and the whole personnel of the Department of Traumatology  and
Orthopedics of the 
Central Clinical Hospital of WAM, Warsaw, for their hospitality and care   
while this write-up
has been prepared.

\end{document}